\documentclass[aps,pre,twocolumn]{revtex4}
\usepackage{graphics}

\begin{document}


\preprint{RIT-01}
\addtolength{\topmargin}{1in}

\title{Two-dimensional Packing in Prolate Granular Materials}


\author{K.~Stokely}
\author{A.~Diacou}
\author{Scott~V.~Franklin}
\email[]{svfsps@rit.edu}
\homepage[]{http://piggy.rit.edu/franklin/}
\affiliation{Dept. of Physics, Rochester Institute of Technology}

\date{\today}

\begin{abstract}
We investigate the two-dimensional packing of extremely prolate
(aspect ratio $\alpha=L/D>10$) granular materials, comparing
experiments with Monte-Carlo simulations.  In experimental piles of
particles with aspect ratio $\alpha=12$ we find the average packing
fraction to be $0.68\pm0.03$.  Both experimental and simulated piles
contain a large number of horizontal particles, and particle alignment
is quantified by an orientational order correlation function.  In both
simulation and experiment the correlation between particle orientation
decays after a distance of two particle lengths.  It is possible to
identify voids in the pile with sizes ranging over two orders of
magnitude.  The experimental void distribution function is a power law
with exponent $-\beta=-2.37\pm0.05$.  Void distributions in simulated
piles do not decay as a power law, but do show a broad tail.  We
extend the simulation to investigate the scaling at very large aspect
ratios.  A geometric argument predicts the pile number density to
scale as $\alpha^{-2}$.  Simulations do indeed scale this way, but
particle alignment complicates the picture, and the actual number
densities are quite a bit larger than predicted.
\end{abstract}


\maketitle
\section{Introduction}
One of the more striking features of piles of very prolate granular
materials (large aspect ratio $\alpha=L/D$) is the connected network
that forms at comparatively low packing fractions.  The formation of
this network is often commercially undesirable.  LCD screens, for
example, cannot function if the molecules are entangled, and lumber
floating down a river stops when the logs jam.  There are, however,
practical applications for such ``jammed'' networks.  Piles of large
aspect-ratio materials are extremely rigid, even at low packing
fractions, and have a high strength:weight ratio.  At the extremely
small scale, networks of carbon nanotubes are a possible mechanism for
conducting energy to and from nano-devices~\cite{Raffaelle}.

\begin{figure}
\scalebox{0.5}{\includegraphics{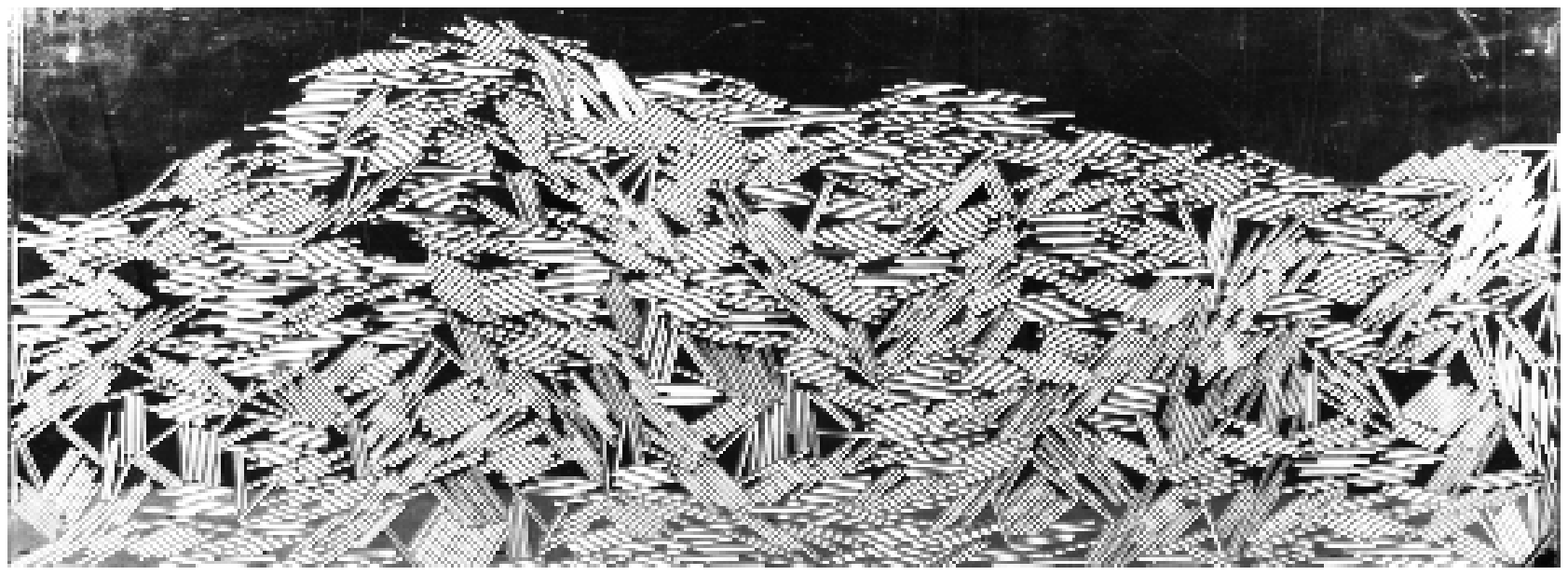}}
\vskip 0.1in
\scalebox{0.40}{{\includegraphics{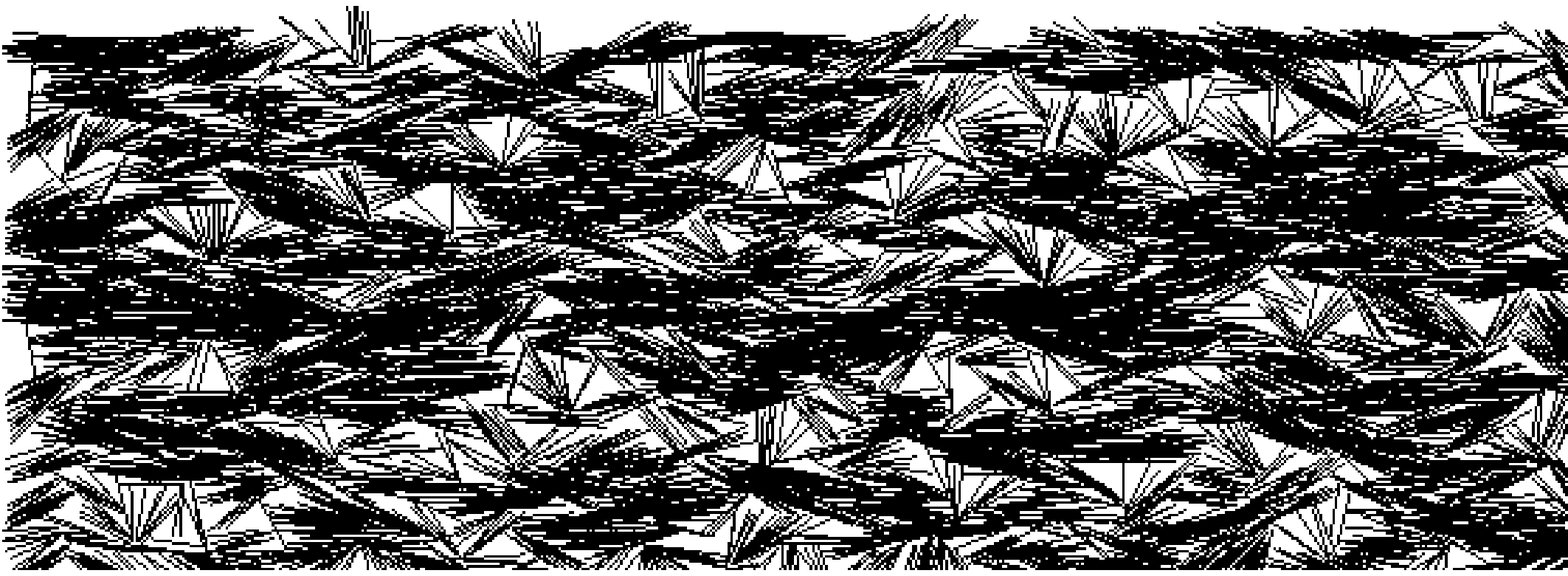}}}
\caption{\label{pile_pics}{\bf Top:} 2-d pile of $\alpha=12$ acrylic
rods backlit with fluorescent lights.  The middle third of each
particle appears bright.  {\bf Bottom:} Simulated pile of $\alpha=12$
particles.  Both piles show particles aligning and a wide distribution
of void sizes.}
\end{figure}

Little is known about even basic characteristics of piles formed from
rod-like particles, most research on non-spherical particles, whether
in two\cite{Buch_Brad_1,Buch_Brad_2,Mustoe1,Mustoe2,Cleary,Rankenburg}
or three~\cite{Villarruel} dimensions, being limited to $\alpha<5$.
The rigidity of such piles is due to particle entanglement, with
particle rotation extremely constrained.  The statistics of particle
orientations which determine these constraints, however, is not
known.  While it seems obvious that particles will align, in fact
two-dimensional piles contain a number of orthogonal particles that
create large voids which dominate the pile landscape (see
Fig.~\ref{pile_pics}).  The only work above $\alpha\sim10$ we are
aware of is that of Philipse~\cite{Philipse1,Philipse2}, who formed
three-dimensional piles of copper wire of aspect ratios ranging from 5
to 77 and explained the $1/\alpha$ scaling of the volume fraction with
a simple geometric model.  As the particles' aspect ratio increased,
they could no longer be poured from their initial container, and
tended to fall out as a solid ``plug''.  The cause of this transition,
which occurs at $\alpha\sim35$, is not known.

\section{Experiment and Simulation}
\begin{figure}
\includegraphics{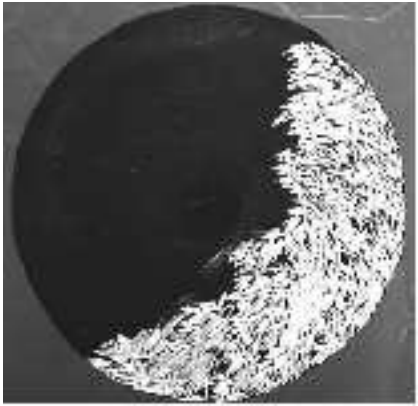}
\caption{\label{wheel}2-dimensional pile of $\alpha=10$ particles showing
an angle of repose of $90^\circ$.  We believe that the ability to
support angles of $90^\circ$ or greater is analogous with Philipse's
observation of solid plugs in three dimensions.}
\end{figure}

To form prolate particles, acrylic rods (diameter $D=0.16$ cm) were
cut to a length $L=1.9$ cm ($\alpha\approx12$) and constrained between
two Plexiglas sheets separated by a spacer 1.25 particle diameters
thick.  The uniform spacing throughout the plates prevents particles
from overlapping; piles are effectively 2-dimensional.  Thicker
spacers result in overlapping particles which are pinched between the
plates and immobile.  Whereas Philipse observed a qualitative increase
in pile rigidity in three dimensions to occur at about an aspect ratio
of 35, we believe that in two dimensions this occurs at much lower
aspect ratios.  We have observed 2-d piles of particles with aspect
ratio 10 to have stable angles of repose of $90^\circ$ or greater (see
Fig.~\ref{wheel}).  We associate this behavior in 2-d with Philipse's
solid plug observed in 3-d.

Piles are formed by distributing particles at random on one Plexiglas
sheet, attaching the second sheet, and slowly rotating the system to
vertical.  The initial distribution is not truly random; local
orientational correlations exist as neighboring particles are forced
to be aligned (or else they would overlap).  This could be avoided in
principle by reducing the number of particles on the plate; in practice,
however, this would be prohibitively slow.  Additional particles are
prepared in a similar manner in an identical cell which is used as a
funnel to pour particles onto the pile.  Piles formed this way are 53
cm wide, typically 25 cm high and contain about 2000 particles.  A
picture of a pile is shown in Fig.~\ref{pile_pics} (top).  The piles
are backlit with fluorescent lights.  The cylindrical rods act as
lenses, displaying a thin bright line throughout the middle of each
particle.  We have written software that identifies connected, collinear
bright pixels in a picture and extracts the particle location and
orientation.  Data reported involve averaging over 19 separate piles.
Despite the less-than-ideal preparation, piles were statistically
consistent; packing fractions varied by $\sim5\%$ and void
distribution and orientational order functions were similarly
reproducible.

Buchalter and Bradley\cite{Buch_Brad_1,Buch_Brad_2} developed a
Monte-Carlo simulation for ellipsoidal particles.  We adapted this for
cylindrical particles and extended the aspect ratio by two orders of
magnitude ($\alpha_{\rm max}=1000$).  Particles move along the nodes
of a discrete lattice ($N\times N$, with $N\sim10L$) and can rotate
freely.  Particles are initially placed at random locations on the
lattice and given random orientations (never being allowed to overlap
with other particles).  A single particle is chosen at random and
moved along a randomly generated displacement/rotation path.  The only
constraint on the motion is that particles cannot move upwards or
overlap with other particles.  The maximum possible distance a
particle can move in one attempt is typically $L/6$.  If an
intersection occurs, the particle is placed at its last allowable
position and a new particle is then chosen for an attempted move.  At
any given time, only one particle is in motion.  The process repeats
until the potential energy (the sum of the particle heights) remains
constant for 5000 time steps, each particle unable to move for, on
average, 10 attempted moves.  A new group of particles is then placed
above the formed pile and allowed to settle.  All piles are at least 7
particle lengths high and we have checked to ensure that additional
pourings do not appreciably change the pile's statistics.  A sample
pile is shown in Fig.~\ref{pile_pics} (bottom).  The length of the
particles, constant through any one pile, is varied from 10 to 1000.
Results for a given aspect ratio are averaged over five piles;
additional piles do not change the statistics.  Additional details
about the simulation's validity, including a discussion comparing the
simulation in the limit as the aspect ratio goes to 1 with
experimental findings, can be found in \cite{Buch_Brad_1}.

\section{Results}
\subsection{Global Pile Characteristics}
The range of packing fractions achievable with 2-d disks under
gravitational forces is quite narrow.  The upper and lower limits are
given by hexagonal ($\phi_{hcp}=\pi/(2\sqrt{3})\approx 0.907$) and
orthgonal ($\phi_{ocp}=\pi/4\approx 0.785$) close packing
respectively, with a random close packing vale of
$\phi_{rcp}\approx0.82$\cite{Howell}.  We find the average packing
fraction of rods with $\alpha=12$ to be $\phi=0.68 \pm 0.03$.  The
lowest measured value was 0.63; the largest 0.72.

The orientational order parameter $Q=\langle \cos(2\theta_i)\rangle$
is used to characterize the angular distribution of particles.
$\theta_i$ is the angle with respect to the horizontal of the $i$-th
particle and the average is over all particles.  $Q$ takes values
ranging from $-1$ (all vertical) to $+1$ (all horizontal), with $Q=0$
indicating an isotropic distribution of angles (or all angles equal to
$45^\circ$).  For the experimental piles $Q=0.33\pm0.05$; simulated
piles have similar values regardless of aspect ratio.

\begin{figure}
\rotatebox{-90}{\scalebox{0.35}{\includegraphics{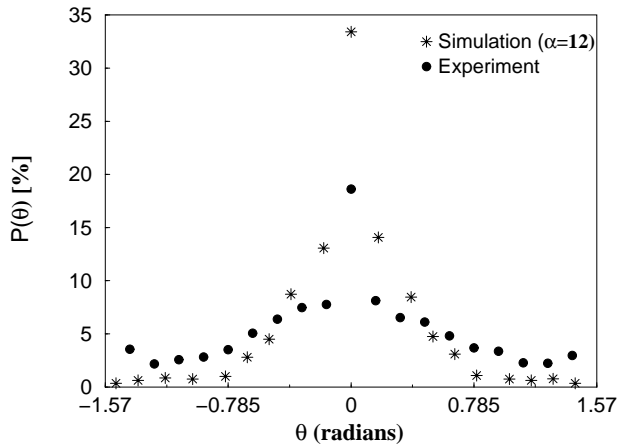}}}
\caption{\label{ang_dist}Angular distribution of experimental
($\bullet$) and simulated ($\ast$) particles.  Both show a peak about
horizontal alignment ($\theta=0$), the experiment having the broader
distribution.}
\end{figure}

\begin{figure}
\scalebox{0.35}{\includegraphics{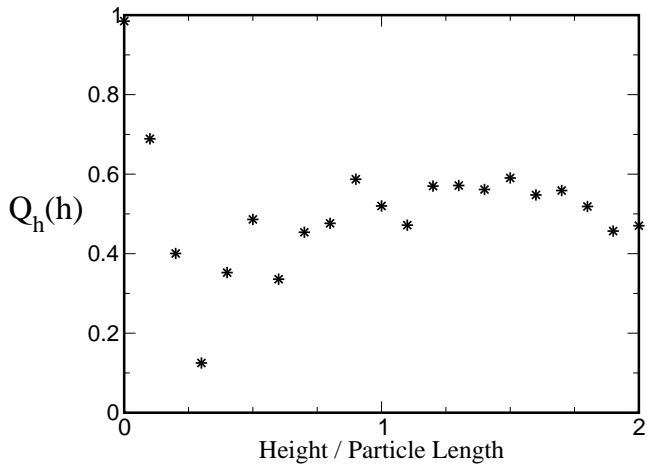}}
\caption{\label{Q_vs_ht}Orientational order parameter $Q_h(h)$ as
function of height in pile.  The reaching of an asymptotic value so
quickly (for $h\sim L$) indicates that the bottom boundary is not
significantly influencing the orientation of particles high up in the pile.
Simulations, which do not have a horizontal bottom boundary but rather
allow the bottom particles to assume any angle, show a similar average
$Q_h(h)$ at all heights.}
\end{figure}

Figure \ref{ang_dist} shows the distribution of particle angles in
experimental and simulated piles.  Both have a peak around $\theta=0$,
indicating a preference for horizontal orientation; this preference is
stronger in the simulated pile.
To determine whether this ordering was caused by the flat bottom
boundary, we calculated the orientational order parameter for all particles
whose centers of mass are at height $h$.  We denote this height
dependent value as $Q_h(h)$.  A plot of $Q_h(h)$ vs. height for
experimental piles is shown in Fig.~\ref{Q_vs_ht}.  $Q_h(h=0)$ is 1,
as particles on the floor must be horizontal.  At a height of $h=L$,
however, $Q_h(h)$ has already decayed significantly.  For heights
greater than a particle length, $Q_h(h)$ fluctuates about an average
value.  From this we infer that the bottom boundary's influence on
particle orientation does not extend beyond one particle length.
Simulations show a similar asymptotic value for $Q_h(h)$, and it
should be noted that the bottom boundary in the simulations does not
impose a horizontal angle on the bottom particles.  Therefore we
believe the tendency for particles to be horizontal, more pronounced the
simulation than in the experiment, is a result of gravity (in the
simulation imposed by the restriction that particles cannot move upwards)
rather than a boundary condition.
Experimental piles also have more vertical particles than the simulation, a
consequence, we believe, of friction between particles which is not
incorporated in the simulation.
			       
\subsection{Distribution of Voids}
\begin{figure}
\rotatebox{-90}{\scalebox{0.35}{\includegraphics{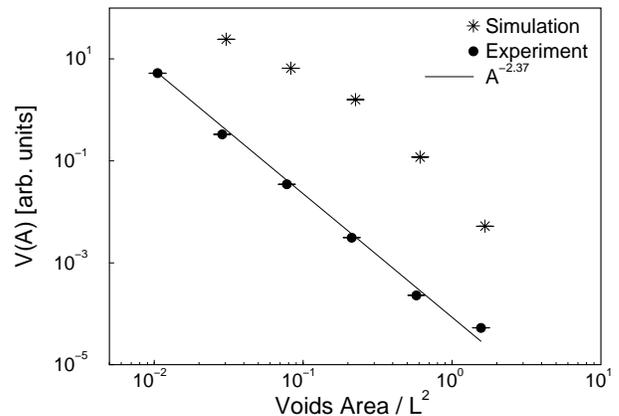}}}
\caption{\label{voids}Void distribution function $V(A)$ as a function
of void area scaled by particle length $L$ squared.  $V(A)$ is
expressed as a percentage of the total number of voids.  Voids in both
experiment ($\bullet$) and simulation ($\ast$, with $\alpha=20$) decay
as a $A^{-\beta}$ with $\beta=2.37\pm0.05$.  This means that smaller
voids occupy a greater total area than the larger, rarer, voids.}
\end{figure}

The appearance of both simulated and experimental piles are dominated
by large, but rare, voids.  From the images we find the number of
voids as a function of void area $A$.  The void distribution functions
$V(A)$ from experimental ($\bullet$) and simulated ($\ast$) piles are
plotted vs. void area $A$ in Fig.~\ref{voids}.  As Fig.~\ref{voids}
shows, experimental void sizes vary by over two decades.  The
experimental data are well-fit by a power law $A^{-2.37\pm 0.05}$
(straight line in Fig.~\ref{voids}).
Simulated piles show a similar decay, although the function does not
seem to follow a power law.  We did not notice any significant
dependence of the void distribution function on the particle aspect
ratio in simulated piles, although this warrants further study.

There are several intriguing consequences to the fact that the
exponent in $V(A)$ is between -2 and -3.  First, the total area taken
up by all voids with size $A$ is $AV(A) \propto A^{-1.37}$. As
$\lim_{A\rightarrow \infty} [AV(A)]\rightarrow 0$, the cumulative
effect of the smaller voids to the pile's area is actually larger than
that of the larger voids.  We also note that the total area occupied
by all voids $\int{AV(A){\rm d}A}$ remains finite as, in fact, it must
for realistic piles.  The total area occupied by voids is related to
the packing fraction $\phi$ by the relation
$$\int{AV(A){\rm d}A} = \int{A^{-1.37}{\rm d}A}=A_{\rm tot}(1-\phi),$$
where $A_{\rm tot}$ is the total pile area.  This relation can be used
to explore the lower limits for the void size, without which the
integral on the left diverges.  This is the subject of current
research.  We also note that the absence of an upper limit for void
size results in the divergence of the integral for the mean square
void area $\langle A^2 \rangle = \int{V(A)A^2{\rm d}A}$.

\subsection{Neighboring Particle Alignment}
\begin{figure}
\hskip 0.3in\rotatebox{-90}{\scalebox{0.35}{\includegraphics{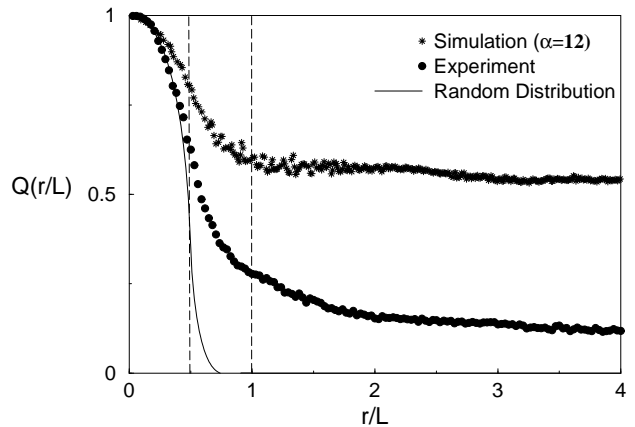}}}
\caption{\label{ang_orient}Orientational correlation function $\tilde{Q}(r/L)$
as a function of center-of-mass separation scaled by particle length.
The solid line represents the correlation resulting from a
distribution where particles assume all allowable angles with equal
probability.  Both experiment and simulation show enhanced alignment
for $r/L$ between 0.5 and 1 (between dashed lines) but noticeably
different asymptotic values due to the different angular distributions
from Fig.~\ref{ang_dist}.}
\end{figure}

Both experiment and simulation find neighboring particles aligning.
This is quantified with an orientational correlation function
$$\tilde{Q}(r)=\langle \cos{(2\Delta \theta_{ij})}\rangle.$$ $\Delta
\theta_{ij}$ is the difference in angle between particles $i$ and $j$
and the average is over all particles whose centers-of-mass separation
is between $r$ and $r+\delta r$.  This function, related to an earlier
orientational order parameter~\cite{Buch_Brad_1,Buch_Brad_2}, takes
values ranging from 1 if particles are parallel to -1 if particles are
perpendicular.  Two particles whose centers-of-mass are quite close
must be aligned and so $\tilde{Q}(r\rightarrow 0) \rightarrow 1$.
Once particle centers-of-mass are separated by more than one particle
length $L$ they can in principle assume any relative orientation and
so $\tilde{Q}(r\rightarrow \infty)\rightarrow 0$.  For comparison, we
calculate analytically the correlation function resulting from a
distribution where particless assume all allowable angles with equal
probability.  This is the simplest possible ordering, the only
constraint being that particles cannot overlap, and is used in simple
geometric models for predicting number density.  The allowable angles
$\theta$ a particle can take with respect to a fixed particle assumed
to lie along the $x$ axis are found as a function of center-of-mass
separation $r$ and angle $\phi$ that the line connecting the
centers-of-mass makes with the $x$ axis.  If the minimum/maximum
allowable angles are given by $\theta_{min}$ and $\theta_{max}$ then
$\tilde{Q}(r)$ is
$$\tilde{Q}(r)=\int_0^{2\pi} {\rm
d}\phi\int_{\theta_{min}(r,\phi)}^{\theta_{max}(r,\phi)}\cos{(2\theta)}{\rm
d}\theta.$$ Figure \ref{ang_orient} shows the $\tilde{Q}(r)$ distribution
resulting from the analytic (line), experimental ($\bullet$), and
simulated ($\ast$) piles.  Both the experimental and simulated piles
show greater correlation between neighboring particles, seen in the
divergence from the analytic line for $r/L>0.5$ (between the dashed
lines), and reach an asymptotic value once particles are separated by more
than two particle lengths.

\begin{figure}
\hskip 1.25in\rotatebox{-90}{\scalebox{0.325}{\includegraphics{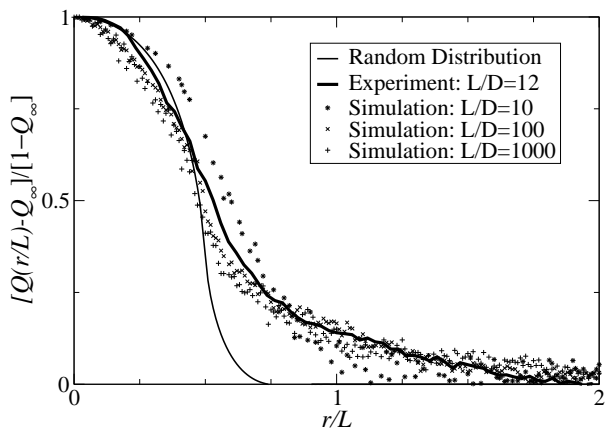}}}
\caption{\label{ang_all_corr}Orientational correlation function
$\tilde{Q}(r/L)$ normalized by the asymptotic value
$\tilde{Q}_{\infty}$ as a function of center-of-mass separation scaled
by particle length.  Simulation results are shown for particle lenghts
varying over three orders of magnitude; the correlation length at
which the asymptotic value is reached is the same for simulations and
experiment.}
\end{figure}

The long-range correlation between particles shown in
Fig.~\ref{ang_orient} does not represent a long-range influence of one
particle on another, but rather results from the overall preference
for particles to be horizontal.  This is confirmed by calculating
$$\tilde{Q}(r\rightarrow \infty)=\int P(\theta) P(\phi) \cos{[2(\theta-\phi)]}
{\rm d}\theta {\rm d}\phi,$$
which assumes the particle angles are drawn at random from the distribution
$P(\theta)$ shown in Fig.~\ref{ang_dist}.  The differences in the
simulation and experimental distribution functions result in
$\tilde{Q}_{\infty}^{\rm exp}=0.16$ and $\tilde{Q}_{\infty}^{\rm sim}=0.53$, agreeing quite well with the asymptotic
values in Fig.~\ref{ang_orient}.
$\tilde{Q}(r/L)$ reaches its asymptotic value for both simulation and
experiment within two particle lengths.  This correlation length is
the same for simulations of particle lengths differing by two orders
of magnitude, as shown in Fig.~\ref{ang_all_corr}. The curves in
Figure~\ref{ang_all_corr} have all been normalized by their asymptotic
value; that is, what is plotted is
$(\tilde{Q}-\tilde{Q}_{\infty})/(1-\tilde{Q}_{\infty})$, which takes
values ranging from 1 to 0.  When thus normalized, all simulated
curves lie very close to the experimental curve.
\subsection{Simulation at Large Aspect Ratios}
\begin{figure}
\hskip
-0.1in\rotatebox{-90}{\scalebox{0.4}{\includegraphics{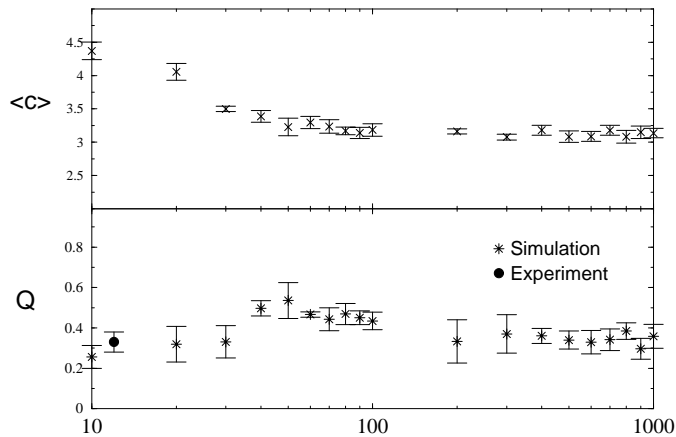}}}
\caption{\label{simul}{\bf Top:} Contact number in simulated piles as
a function of particle aspect ratio.  $\langle c\rangle$ is
independent of aspect ratio for long particles, and reaches a suitable
asymptote, 4.5, comparable withe contact number of disks, as
$\alpha\rightarrow 1$.  {\bf Bottom:} Orientational order parameter
$Q$ as a function of aspect ratio for simulation ($\ast$) and
experiment ($\bullet$).}
\end{figure}

We now extend the simulation to larger aspect ratios and investigate
the scaling of various quantities.  The rigidity of a pile depends on
particles in contact, hence we calculate the {\it contact number}
$\langle c \rangle$.  Fig.~\ref{simul}(top) shows that $\langle c
\rangle$ reaches an asymptotic value of $\approx 3.2$ by aspect ratio
50.  This may seem counter-intuitive, as longer particles in principle
can be in contact with more neighbors.  Orientations that maximize
$\langle c \rangle$, however, are quite rare and the
length-independence of contact number is due to the tendency of
neighboring particles to align and screen one another from other
particles.  As particle aspect ratio decreases, and the particles
become more circular, this screening effect diminishes and the contact
number increases.  Packings of circular disks, for example, show a
contact number between 4 (orthogonal close packing) and 6 (hexagonal
close packing).

Recall that the pile's orientational order is characterized by the
order parameter $Q=\langle \cos(2\theta_i)\rangle$ where $\theta_i$ is
the angle the $i$th particle makes with the horizontal and $Q$ is
averaged over all particles.  As shown in Fig.~\ref{simul}, $Q$ of
simulated piles decreases as the particle length increases reaching an
asymptote of 0.3.  $Q$ for experimental piles of particles with aspect
ratio $\alpha=12$ is $0.33\pm0.05$, comparable with that of of
simulations.

\subsection{Phenomenology}
Philipse gave a simple geometric argument, the Random Contact
Model\cite{Philipse1} (RCM)to explain the low packing fractions of
three-dimensional piles.  We now apply his logic to our
two-dimensional piles and show the discrepancies caused by the
enhanced particle alignment described above.

\begin{figure}
\rotatebox{-90}{\scalebox{0.4}{\includegraphics{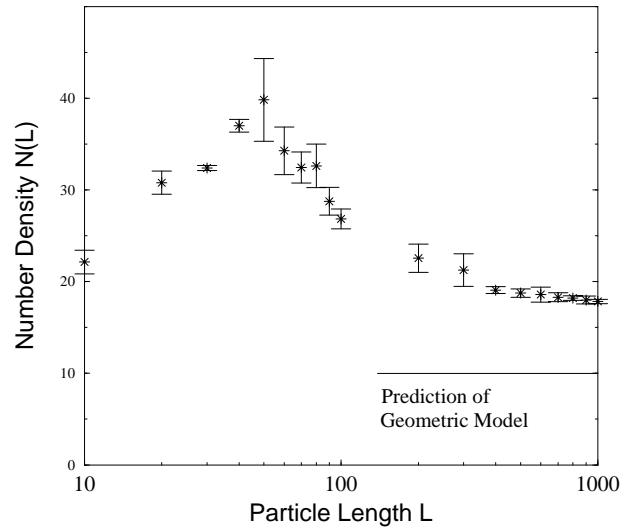}}}
\caption{\label{num_dens} Number density as a function of particle
length.  As predicted by the simple geometric model described in the
text, the number density appears to scale as $L^{-2}$ for large $L$.
The actual densities however, are larger than that predicted by the
model, a consequence of particle alignment.}
\end{figure}

The existence of one particle excludes a fraction of the possible
orientations, called the {\it excluded area} $A_{\rm excl}$, that can
be assumed by a second particle.  If we assume a connected network,
where all particles are in contact with (on average) $\langle
c\rangle$ neighbors and that particles assume all allowable
orientations with equal probability, then the average number density
will be
$$\langle N\rangle=\frac{2 \langle c\rangle}{A_{\rm excl}}.$$
The factor of 2 accounts for the fact that each contact involves two
particles.  We have already shown, however, that the assumption that
contacts are uncorrelated, is not satisfied.  Balberg\cite{Balberg2}
has calculated the excluded area of a stick with length $L$ and width
$W$; to first order $A_{\rm excl}=(2/\pi )L^2$.  With $\langle
c\rangle=3.2$, the number density as a function of aspect ratio
$\alpha$ is predicted to be
$$N(L)=\frac{2\langle c \rangle}{(2L^2/\pi)}=C L^{-2}$$ with $C=10$.
Fig.~\ref{num_dens} shows that $N(L)$ does indeed fall off as $L^{-2}$
for large $L$.  The constant, however, is larger than that
predicted by the RCM (flat line in Fig.~\ref{num_dens}).
Piles are therefore more dense than predicted, implying that the
excluded area of particles is about 33\% {\it less} than that in an
isotropic distribution.  This is a result of particle alignment, seen
earlier in Fig.~\ref{ang_orient}.  We also note that the scaling as
$\alpha^{-2}$ is realized only for the largest of aspect ratios, while
the constancy of contact number occurs much earlier.

\section{Conclusions}
We have presented the first quantitative characterization of
two-dimensional piles formed from prolate ($\alpha>10)$ granular
materials finding, for example, the packing fraction of particles with
aspect ratio $\alpha=12$ to be $0.68\pm0.03$.  Particles separated by
less than two particles lengths show a greater orientational
correlation than would be found in a random pile; particles separated
by more than two lengths are uncorrelated except for the general
preference for horizontal alignment imposed by gravity.  The void
distribution function in experimental piles obeys a power law with
exponent $-\beta=2.37\pm0.05$; Monte-Carlo simulations show similar
angular correlations and void distribution functions.  Simulations
have a greater number of horizontal particles, however, and thus
produce piles with larger number densities than found in both
experiment and simple geometric models.

We thank E.~F.~Redish for first questioning the characteristics of
pickup sticks and John C. Crocker for calling attention to the work of
Philipse.  Eric R. Weeks and L.~S.~Meichle have provided invaluable
advice throughout this project.  Saul Lapidus and Peter Gee were
involved in much of the original setup of the experiment.


\bibliography{rods_references}
\end{document}